\documentclass[conference]{IEEEtran}
\usepackage[latin9]{inputenc}
\usepackage{array}
\usepackage{amsmath}
\usepackage{amssymb}
\usepackage{graphicx}

\makeatletter

\providecommand{\tabularnewline}{\\}

\IEEEoverridecommandlockouts
\usepackage{cite}
\usepackage{amsfonts}
\usepackage{textcomp}
\usepackage{xcolor}
\def\BibTeX{{\rm B\kern-.05em{\sc i\kern-.025em b}\kern-.08em
    T\kern-.1667em\lower.7ex\hbox{E}\kern-.125emX}}

\usepackage[linesnumbered,ruled,vlined]{algorithm2e}
\usepackage{algpseudocode}

\algnewcommand\algorithmicforeach{\textbf{for each}}
\algdef{S}[FOR]{ForEach}[1]{\algorithmicforeach\ #1\ \algorithmicdo}

\usepackage{pgf}
\usepackage{pgfplots}

\makeatother

\begin{document}
\title{Scaling up the self-optimization model by means of on-the-fly computation
of weights}
\author{\IEEEauthorblockN{Natalya Weber} \IEEEauthorblockA{\textit{Embodied Cognitive Science Unit} \\
\textit{\emph{Okinawa Institute of Science and }}\\
\textit{\emph{Technology Graduate University}}\\
Okinawa, Japan\\
 natalya.weber@oist.jp} \and \IEEEauthorblockN{Werner Koch} \IEEEauthorblockA{\textit{Independent Scholar}\\
Dresden, Germany \\
ORCID: 0000-0001-7246-0434}\and\IEEEauthorblockN{Tom Froese} \IEEEauthorblockA{\textit{Embodied Cognitive Science Unit} \\
\textit{\emph{Okinawa Institute of Science and }}\\
\textit{\emph{Technology Graduate University}}\\
Okinawa, Japan\\
tom.froese@oist.jp}}
\maketitle
\begin{abstract}
The Self-Optimization (SO) model is a useful computational model for
investigating self-organization in ``soft'' Artificial life (ALife)
as it has been shown to be general enough to model various complex
adaptive systems. So far, existing work has been done on relatively
small network sizes, precluding the investigation of novel phenomena
that might emerge from the complexity arising from large numbers of
nodes interacting in interconnected networks. This work introduces
a novel implementation of the SO model that scales as $\mathcal{O}\left(N^{2}\right)$
with respect to the number of nodes $N$, and demonstrates the applicability
of the SO model to networks with system sizes several orders of magnitude
higher than previously was investigated. Removing the prohibitive
computational cost of the naive $\mathcal{O}\left(N^{3}\right)$ algorithm,
our on-the-fly computation paves the way for investigating substantially
larger system sizes, allowing for more variety and complexity in future
studies. 
\end{abstract}


\IEEEoverridecommandlockouts
\IEEEpubid{\begin{minipage}{\textwidth}\ \\[12pt]\copyright\ 2022 IEEE. Personal use of this material is permitted. Permission from IEEE must be obtained for all other uses, in any current
or future media, including reprinting/republishing this material for advertising or promotional purposes, creating new collective works, 
for resale or redistribution to servers or lists, or reuse of any copyrighted component of this work in other works.\end{minipage}}
\IEEEpubidadjcol

\section{Introduction\label{sec:Introduction}}

Self-organization is the process by which stable structures, or patterns,
emerge spontaneously as a result of nonlinear interaction between
a large number of components. Self-organization is observed in all
matter on any scale, from the nanoscale of quantum particles to the
macroscale of galaxies \cite{jooss_self-organization_2020}, and in
particular life forms are known to be highly self-organized dynamic
molecular systems \cite{de_la_fuente_self-organization_2021,kelso_dynamic_1995}.
Consequently, self-organization has played a major role in soft (simulated),
hard (robotic), and wet (chemical and biochemical) domains of Artificial
life (ALife)\cite{gershenson_self-organization_2020}. Adaptive networks
are networks whose states interact with each other, are highly coupled,
and keep changing due to the system\textquoteright s own dynamics,
producing emergent behavior that would not be seen in other forms
of networks\cite{sayama_modeling_2013}. As such, adaptive networks
have been used to describe the self-organization of various systems,
ranging from systems like the brain or the nervous system (due to
the interaction between neurons that produces behavioral and cognitive
patterns\cite{kelso_dynamic_1995}), to social and engineered complex
systems\cite{sayama_modeling_2013}. 

Of particular interest for the current paper is the Self-Optimization\footnote{Originally termed as \emph{self-modeling} framework \cite{watson_optimization_2011},
here we adopt the terminology later proposed in \cite{zarco_self-optimization_2018}.} (SO) model \cite{watson_effect_2009,watson_optimization_2011}. The
SO model demonstrates how a system (an adaptive network) that augments
its behavior with an associative memory of its own attractors is capable
of solving constraint optimization problems by spontaneous (without
an external reward signal) self-organization. The SO model has been
shown to be general enough to model various complex adaptive systems,
from gene regulation networks\cite{watson_associative_2010} and nematode
worms \cite{morales_self-optimization_2019,morales_unsupervised_2020},
to selfish agents\cite{watson_global_2011} and sociopolitical networks\cite{Froese2014,froese_modeling_2018}
(more references are given in Table \ref{tab:Literature-review}). 

Aside of its domain-generality, what makes the SO model particularly
interesting is that it is simple to implement and understand, yet
with sufficient complexity to exhibit a multitude of effects relevant,
evidently, for various fields. It remains to be answered, however,
how practical is the model? The original model \cite{watson_effect_2009,watson_optimization_2011}
was used on networks with non-directed connections constrained to
symmetric weights with a discrete state. Since then several studies
showed that the SO model can be generalized to asymmetric weight matrices
and continuous-state\cite{zarco_self-modeling_2018}, multiple directed
connections\cite{morales_self-optimization_2019}, and continuous-time\cite{zarco_self-optimization_2018}. 

\IEEEpubidadjcol

In this work, we continue this effort of generalization by showing
that the SO model is also scalable to large systems, which is crucial
to support arguments in the literature that the model can provide
insights into biological and social systems that consist of thousands,
tens of thousands, or more elements. We do so by deriving a novel
implementation that reduces the scaling from $\mathcal{O}\left(N^{3}\right)$
of the straightforward implementation of the mathematical model to
$\mathcal{O}\left(N^{2}\right)$ through rearranging the terms. Consequently,
an analysis of networks with sizes that would otherwise have incurred
prohibitive computational requirements becomes accessible.

The rest of the paper is structured as follows: Section \ref{sec:Background}
introduces the model and its regular straightforward computational
implementation. Section \ref{sec:Implementation} describes the proposed
novel implementation. In Section \ref{sec:Results} we will present
computation results for our work and in Section \ref{sec:Conclusions}
we will draw the conclusions of this work.

\section{Background\label{sec:Background}}

\subsection{Theoretical model\label{subsec:Theoretical-model}}

The self-optimization (SO) model is a model introduced by \cite{watson_effect_2009,watson_optimization_2011},
that combines two well-known dynamical behaviors of recurrent neural
networks: \emph{constraint satisfaction}, that can be understood as
an energy minimization process, and \emph{autoassociative memory}
that forms as a result of associatively modifiable recurrent\textbackslash{}
connections in the network. This combination results in a significant
improvement in the ability of the neural network to find configurations
that satisfy constraints, and thus performs effective optimization.

At the basis of the model is a Hopfield neural network\cite{hopfield_neural_1982}.
The Hopfield network is a representation of a physical system with
$N$ nodes (or ``neurons''), where each node can be in one of two
possible states, either $s=1$ or $s=-1$. At at any given time, the
state of the system is defined by a vector $\mathbf{S}\left(t\right)=\left\{ s_{1}\left(t\right),...,s_{N}\left(t\right)\right\} $.
Overall there are $2^{N}$ possible states of the system. The connections
between the nodes are defined by a weight matrix, $\mathbf{W}$, of
size $N\times N$, where $w_{ij}=0$ means that there is no connection
present between nodes $i$ and $j$. The dynamics of the system is
chosen to be asynchronous, that is at each time step $t$, a node
$i$ is chosen at random and updated using the following rule
\begin{equation}
s_{i}\left(t+1\right)=\theta\left[\sum_{j}^{N}w_{ij}s_{j}\left(t\right)\right],\label{eq:state_update}
\end{equation}
where $w_{ij}$ are elements of the weight matrix $\mathbf{W}$, and
$\theta$ is a Heaviside threshold function (taking values -1 for
negative arguments and +1 otherwise). When the connection strengths
are symmetric (i.e. $w_{ij}=w_{ji}$), the Hopfield model admits a
Hamiltonian description, and an energy function can be defined
\begin{equation}
E_{\mathbf{W}}\left(t\right)=-\frac{1}{2}\sum_{ij}^{N}w_{ij}s_{i}\left(t\right)s_{j}\left(t\right).\label{eq:E}
\end{equation}

\begin{figure}     
\begin{center}  
\input{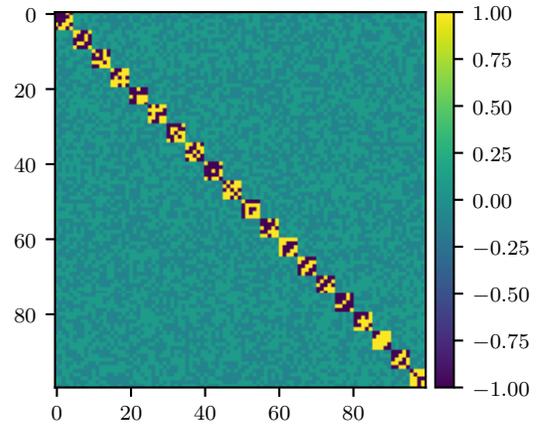}    
\end{center}     
\caption{A symmetric modular connectivity weight matrix for a system of size \(N=100\) with 20 modules of size \(k=5\), intramodule weights set at random to either 1 or -1, and intermodule weights set at random to either 0.1 or -0.1. For details on various constraint problems see {\cite{watson_transformations_2011}}.}
\label{fig:W}
\end{figure}

\begin{figure}     
\begin{center}  
\input{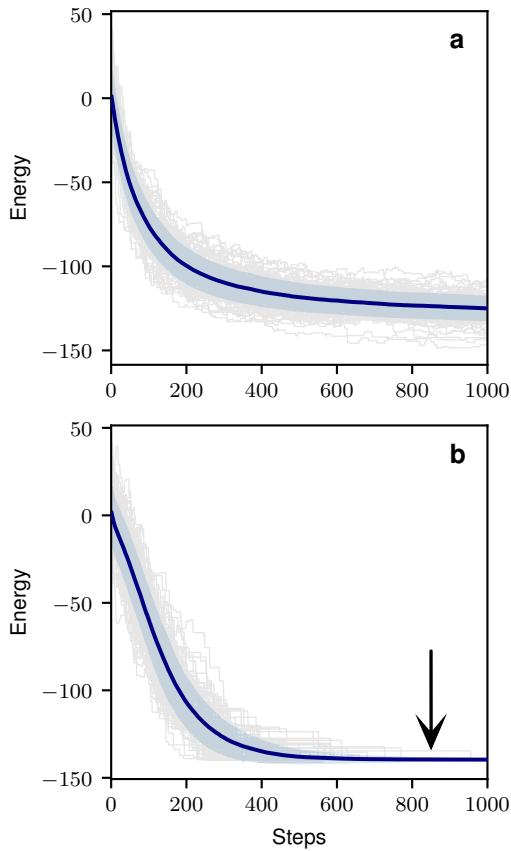}    
\end{center}     
\caption{Dynamics of the SO model. (a) Without learning. This is equivalent to the dynamics of a regular Hopfield network, where the system will converge to various minima according to the initial state. (b) With learning, \eqref{eq:w_update}, \(\alpha=1\times10^{-6}\). On both plots, the energy is computed using \eqref{eq:E} for the chosen initial weight matrix in Fig.\,\ref{fig:W}, and the state at each step is updated asynchronously according to \eqref{eq:state_update}. The plots show the  individual traces of energy for 100 different initial random states in grey, the energy mean for 1000 different initial random states in dark blue, and its standard deviation in light blue.}
\label{fig:E_conv}
\end{figure}

Given a fixed weight matrix $\mathbf{W}$ that represents the constraints
of a constraint optimization problem (e.g. in Fig.~\ref{fig:W}),
a random initial state $\mathbf{S}$, and the dynamics described by
update rule \eqref{eq:state_update}, the system will converge towards
some local minimum in the attractor landscape described by \eqref{eq:E},
which are locally optimal solutions to that problem given by the weight
matrix $\mathbf{W}$. If the system is initialized to another random
state, it will again converge, but perhaps to another local minimum
as shown in Fig.~\ref{fig:E_conv}a. 

The learning phase consists of updating the weights according to an
associative rule in a form of Hebbian learning \cite{hebb_organization_1949}
\begin{equation}
w_{ij}\left(t+1\right)=w_{ij}\left(t\right)+\mathrm{d}w_{ij}\left(t\right),\quad\mathrm{d}w_{ij}=\alpha s_{i}s_{j}\label{eq:w_update}
\end{equation}
where $\alpha>0$ is a learning rate constant. The system is repeatedly
allowed to converge to a local attractor. After performing $T$ steps
the state is reset to a new random state. In total the procedure is
run for $R$ resets. The energy of the system is always computed with
regards to the initial weight matrix, but the state update depends
on the learned weight matrix.

Given a suitable parameter $\alpha$, the system will go through a
self-organization such that regardless of the initial state of the
system, it converges to a lower energy state (a global attractor)
as shown by the arrow in Fig.~\ref{fig:E_conv}b, despite never having
visited that attractor before.

\begin{figure}     
\begin{center}  
\input{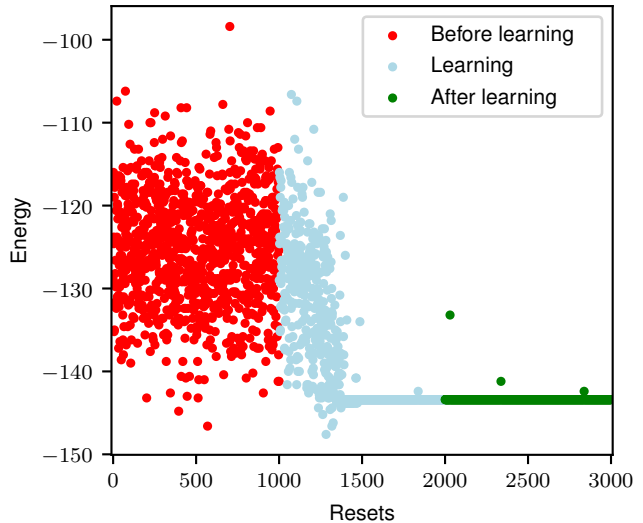}    
\end{center}     
\caption{Distribution of the attractor energies with and without learning for the chosen initial weight matrix in Fig.\,\ref{fig:W}. The points represent the energy at the end of convergence (the energies at step \(t=1000\) in Fig.\,\ref{fig:E_conv}), for a set without learning (resets 1-1000, red), during learning (1001-2000, blue), and after learning (2001-3000, green).}
\label{fig:Distribution-E}
\end{figure}

When we change the perspective on convergence of the state under the
update rule \eqref{eq:state_update}, as illustrated in Fig.~\ref{fig:E_conv},
and instead we plot only the energy at the end of convergence for
a set of random initial states it allows us to look at the distribution
of the attractor energies with and without learning (Fig.~\ref{fig:Distribution-E}).

The repeated energy minimization and slow accumulation of changes
to the connections modifies the network to constitute an associative
memory of its own attractor states. But in this case, the accumulation
of weights also changes the sizes of basins of attraction, such that
some attractor basins are enlarged at the expense of others, eventually
resulting in a positive feedback that significantly improves the system's
ability to find configurations that resolve the constraints of the
original weight matrix to a globally minimal energy state \cite{watson_optimization_2011}. 

In next section we discuss the computational implementation of the
SO model.

\subsection{Computational implementation}

A review of the SO literature shows that large networks are generally
not studied (Table \ref{tab:Literature-review}). Given the possibility
for novel phenomena to emerge from the complexity possible when large
numbers of nodes interact in interconnected networks, it is pertinent
to ask why that limitation to relatively small network sizes persists?
Since the implementation details in these articles are usually not
discussed\footnote{Among all the references in Table \ref{tab:Literature-review}, only
\cite{shpurov_combining_2021} provided the code, and according to
it they perform the straightforward computation of \eqref{eq:w_update}.}, it is fair to assume that their authors implement a straightforward
direct computation as shown in Algorithm \ref{Alg-1}. 

\begin{table}
\noindent\begin{minipage}[t]{1\columnwidth}%
\caption{Literature review of system sizes used with the Self-Optimization
model\label{tab:Literature-review}}

\begin{center}
\begin{tabular}{|c|>{\centering}p{6cm}|}
\hline 
N & Reference\tabularnewline
\hline 
\hline 
30 & Watson et al. (2010) \cite{watson_associative_2010}\footnote{Continuous state}\tabularnewline
\hline 
36 & Mills et al. (2011) \cite{mills_symbiosis_2011}\tabularnewline
\hline 
66 & Froese et al. (2014) \cite{Froese2014}\tabularnewline
\hline 
70 & Froese et al. (2018) \cite{froese_modeling_2018}\tabularnewline
\hline 
10,36,100 & Woodward et al. (2015) \cite{woodward_neural_2015}\emph{}\textsuperscript{\emph{a}}\tabularnewline
\hline 
100 & Watson et al. (2009) \cite{watson_effect_2009}, Davies et al. (2011)
\cite{davies_if_2011}, Zarco et al. (2018) \cite{zarco_self-modeling_2018}\emph{}\textsuperscript{\emph{a}}\tabularnewline
\hline 
100,120 & Watson et al. (2011) \cite{watson_global_2011}\tabularnewline
\hline 
30,40,100,120 & Mills et al. (2011) \cite{mills_emergent_2011}\tabularnewline
\hline 
32,128 & Fernando et al. (2010) \cite{fernando_neuronal_2010}\tabularnewline
\hline 
150 & Watson et al. (2011) \cite{watson_optimization_2011}\tabularnewline
\hline 
200 & Watson et al. (2011) \cite{watson_transformations_2011}\tabularnewline
\hline 
279 & Morales et al. (2020) \cite{morales_unsupervised_2020}\tabularnewline
\hline 
280 & Morales et al. (2020) \cite{morales_distribution_2020}\tabularnewline
\hline 
282 & Morales et al. (2019) \cite{morales_self-optimization_2019}\tabularnewline
\hline 
400 & Power et al. (2015) \cite{power_what_2015}, Shpurov et al. (2021)\cite{shpurov_combining_2021}\tabularnewline
\hline 
\end{tabular}
\par\end{center}%
\end{minipage}
\end{table}

\begin{algorithm}
\caption{Direct implementation of learning stage in the SO model}\label{Alg-1} 
\SetAlgoLined
\DontPrintSemicolon

\For{$ 0<r\leq R $}{
\texttt{Generate random state} $\mathbf S$\;
\For{$ 0<t\leq T $}{
\texttt{Choose random index} $0<i_{t}\leq N$\;
\texttt{Compute binary state} $s_{i_{t}}$ \Comment{Eq.\,(1)}\;
\texttt{Compute} $\mathrm d \mathbf W$ \texttt{and} $\mathbf W$ \Comment{Eq.\,(3)}\;
}
}
\end{algorithm}

This is fine, however, much of the computational effort in Algorithm
\ref{Alg-1} is dominated by the need to update the weight matrix
$\mathbf{W}$ at each learning step according to \eqref{eq:w_update}
which subsequently is required for the state update \eqref{eq:state_update}.
Since the weight matrix has $N^{2}$ elements and there are $\propto N$
steps per reset, the weight matrix update scales as $\mathcal{O}\left(N^{3}\right)$.
With a simple addition to be performed, it is primarily memory bandwidth
constrained. This significantly limits the size one can investigate
in a research project. 

In next section we present a novel implementation that reduces this
dependence on the size of the system to $\mathcal{O}\left(N^{2}\right)$.

\section{Novel implementation\label{sec:Implementation}}

Whether the SO model is scalable or not is a really important consideration
that has not been addressed by previous work. However, a straightforward
computation of \eqref{eq:w_update} and \eqref{eq:E} takes a very
long time to compute for large systems because of the $\mathcal{O}\left(N^{3}\right)$
dependence on the size of the system. 

In order to test the effect of scaling the SO model to substantially
higher node counts, a series of mathematically transparent but algorithmically
significant manipulations were implemented to reduce the time of computation.
Here we discuss the dominant reduction of memory accesses from $\mathcal{O}\left(N^{3}\right)$
to $\mathcal{O}\left(N^{2}\right)$, other minor manipulations are
outlined in the Appendix. 

The cost of the memory bandwidth constraint can be reduced dramatically
based on the observation that for large $N$, the change of the weight
matrix $d\mathbf{W}$ is mostly constant from step to step with (on
average) only a few changes according to \eqref{eq:state_update}.
If the history of state changes $\mathbf{S}\left(t\right)$ according
to \eqref{eq:state_update} is recorded, the weight matrix need never
be updated explicitly but rather the value of $\mathbf{W}\left[i,:\right]$
at time $t$ needed for the state update $s_{i}$ in \eqref{eq:state_update}
at time $t$ can be computed on the fly as follows: The row $\mathbf{W}\left[i,:\right]$
is computed as the sum of $\mathbf{W}\left[i,:\right]$ at $t^{'}$
(the last time that state $s_{i}$ was updated), and $\left(t-t^{'}\right)d\mathbf{W}\left[i,:\right]$
at time $t^{'}$. Since this sum ignores the state changes since $t^{'}$,
these changes are then applied through $\left(t-t^{'}\right)$ individual
changes to $\mathbf{W}\left[i,:\right]$ to correct for this omission
(see Algorithm \ref{Alg-2}).

\begin{algorithm}
\caption{"On-the-fly" implementation of learning stage in the SO model}\label{Alg-2} 
\SetAlgoLined
\DontPrintSemicolon

\For{$ 0<r\leq R $}{
\texttt{Generate random state} $\mathbf S$\;
\texttt{Initialize state-to-time map} $\mathbf t'(i=1\ldots N)=1$\;
\texttt{Allocate time-to-state map} $\mathbf i'(t=1\ldots T)$\;
\texttt{Allocate state change history} $\mathbf s'(t=1\ldots T)$\;
\For{$ 0<t\leq T $}{
\texttt{Choose random index} $0<i_{t}\leq N$\;
\texttt{Compute the constant change} $\mathbf W[i_{t},:]=\mathbf W[i_{t},:]+(t-t'(i_{t}))\mathrm d \mathbf W[i_{t},:]$\;
\texttt{Correct for the state changes since} $t'(i_{t}) \texttt{:}$\;
\For{$ t'(i_{t})\leq t''< t $}{
$\mathbf W[i_t,i'(t'')]=\mathbf W[i_t,i'(t'')]+\left(s_{i_{t}}s'(t'')-\mathrm d \mathbf W[i_t,i'(t'')]\right)(t-t'') $\;
}
\texttt{Store the step} $t'(i_{t})=t$\;
\texttt{Store the index} $i'(t)=i_{t}$\;
\texttt{Compute binary state} $s_{i_{t}}$ \Comment{Eq.\,(1)}\;
\texttt{Store the state change} $s'(t)=s_{i_{t}}$\;
}
\texttt{Accumulate remaining corrections at} $t=T$\;
\For{$ 0<i\leq N $}{
\texttt{Compute the constant change} $\mathbf W[i,:]=\mathbf W[i,:]+(t-t'(i))\mathrm d \mathbf W[i,:]$\;
\texttt{Correct for the state changes since} $t'(i)$ \texttt{:}\;
\For{$ t'(i)\leq t''< t $}{
$\mathbf W[i_t,i'(t'')]=\mathbf W[i_t,i'(t'')]+\left(s_{i_{t}}s'(t'')-\mathrm d \mathbf W[i_t,i'(t'')]\right)(t-t'') $\;
}
}
}
\end{algorithm}

This procedure is computationally more elaborate, but never needs
to adjust more than one row of $\mathbf{W}$ for each step. It thus
reduces the memory accesses from $\mathcal{O}\left(N^{3}\right)$
to $\mathcal{O}\left(N^{2}\right)$. In the next section we compare
the execution time between the two procedures numerically. 

\section{Results\label{sec:Results}}

After a proof of principle of the above-mentioned optimized procedure
was attained, the algorithm was implemented as a compiled FORTRAN
module to be loaded from Julia or Python. This combination retains
the flexibility of rapid testing of many different ideas as offered
by an interpreted language while benefiting from the speed of execution
of a compiled language for the most time critical parts. We checked
that the ``on-the-fly'' implementation and the regular direct implementation
return identical states, energies, and weight matrices for a given
simulation seed. 

\subsection{Execution time comparison}

In this section we present the result of applying all the consecutive
changes described in previous section and the Appendix to the code.

To compare between the two procedures we used the BenchmarkTools Julia
package with the minimum as the estimator for the true run time of
a benchmark instead of the mean or median, since it provides the smallest
error \cite{chen_robust_2016}. We first made a comparison for a typical
run of 1000 resets (see Fig.~\ref{fig:Distribution-E} for reference)
between the direct and the on-the-fly implementation. For comparison,
the non-learning stage is also shown.

\begin{figure}     
\begin{center}  
\input{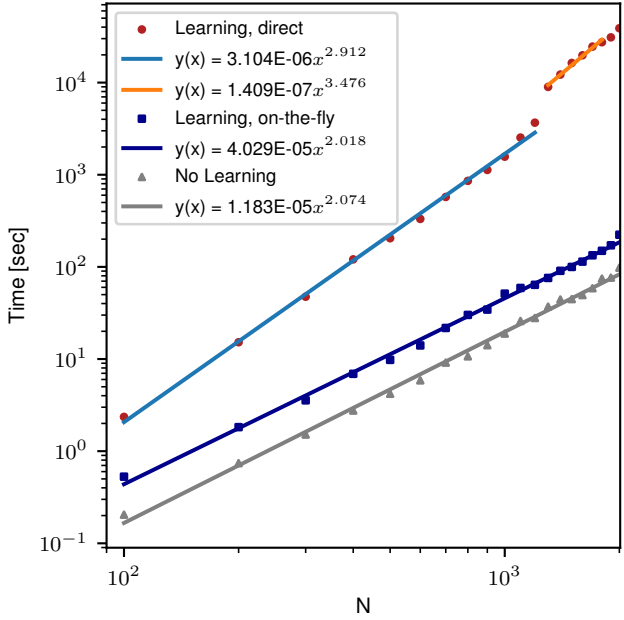}    
\end{center}     
\caption{Execution time comparison of the SO model for regular direct implementation, the on-the-fly implementation, and the non-learning stage for 1000 resets. Six benchmark samples were used for \(N\in[100,1800]\) and five samples for \(N\in[1900,2000]\). Note that around \(N=1200\), there is a jump in the the execution time for the direct implementation. This is likely due to a processor cache limit being exceeded. We did not further investigate this behavior as it does not significantly affect the scaling of the algorithm as evidenced by the two separate fits for \(N<1200\) and \(N>1200\).}
\label{fig:1000resets}
\end{figure}

\begin{figure}     
\begin{center}  
\input{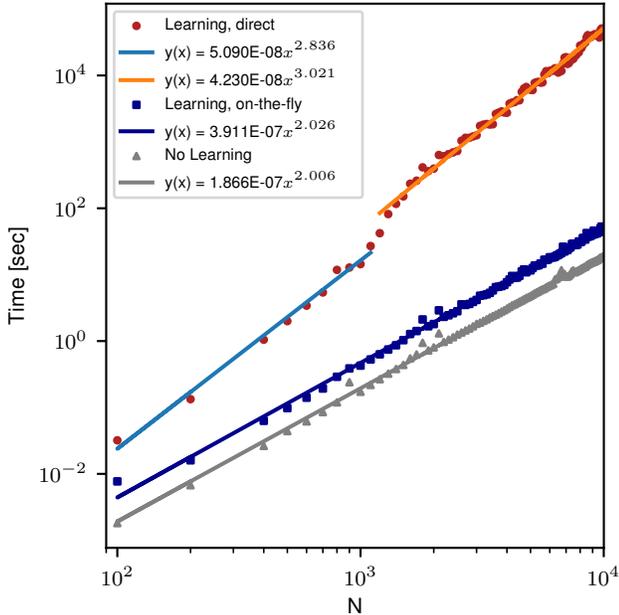}    
\end{center}     
\caption{Execution time comparison of the SO model for regular direct implementation, the on-the-fly implementation, and the non-learning stage for 10 resets. Six benchmark samples were used for \(N\in[100,7100]\) and five samples for \(N\in[7200,10000]\). For an explanation of the jump in the fit for the direct implementation see Fig.\,\ref{fig:1000resets}.}
\label{fig:10resets}
\end{figure}

Figure~\ref{fig:1000resets} shows that, indeed, not only is the
novel implementation faster than the regular one, but it scales differently.
That said, already for $N>2000$ the execution time becomes prohibitive
for continuing our comparison. For this reason, in order to investigate
how the procedures compare for even larger systems we continued the
comparison with a reduced number of 10 resets as shown in Fig.~\ref{fig:10resets}.

Section \ref{subsec:N-10000} shows the results of a simulation for
a system of size $N=10000$. 

\subsection{Simulation results for $N=10000$\label{subsec:N-10000}}

\begin{figure}     
\begin{center}  
\input{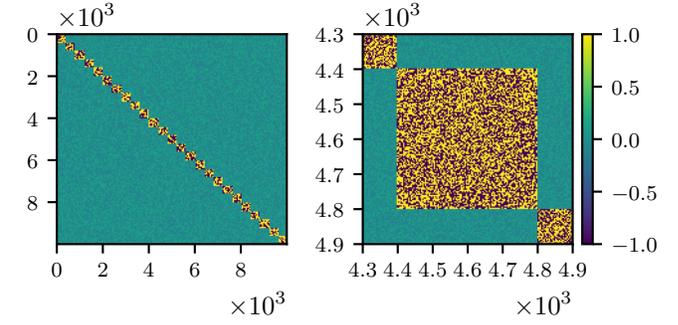}    
\end{center}     
\caption{A symmetric modular connectivity weight matrix for a system of size \(N=10000\) with 25 modules of size \(k=400\), intramodule weights set at random to either 1 or -1, and intermodule weights set at random to either 0.1 or -0.1.}
\label{fig:W10k}
\end{figure}

To investigate the scalability of the SO model, similar computations
to that in Sec.~\ref{subsec:Theoretical-model} were performed for
larger systems and similar behavior was observed. For brevity, we
only show results for $N=10000$. The weight matrix $\mathbf{W}$
(Fig.~\ref{fig:W10k}) has symmetric modular connectivity with modules
of size $k=400$, with intramodule weights set at random to either
1 or -1, and intermodule weights set at random to either 0.1 or -0.1.
Note that the size of the individual modules in this initial large
scale test is equivalent to biggest total system size tested in previous
work! (see Table~\ref{tab:Literature-review}). 

\begin{figure}
\centering{}\includegraphics[scale=0.24]{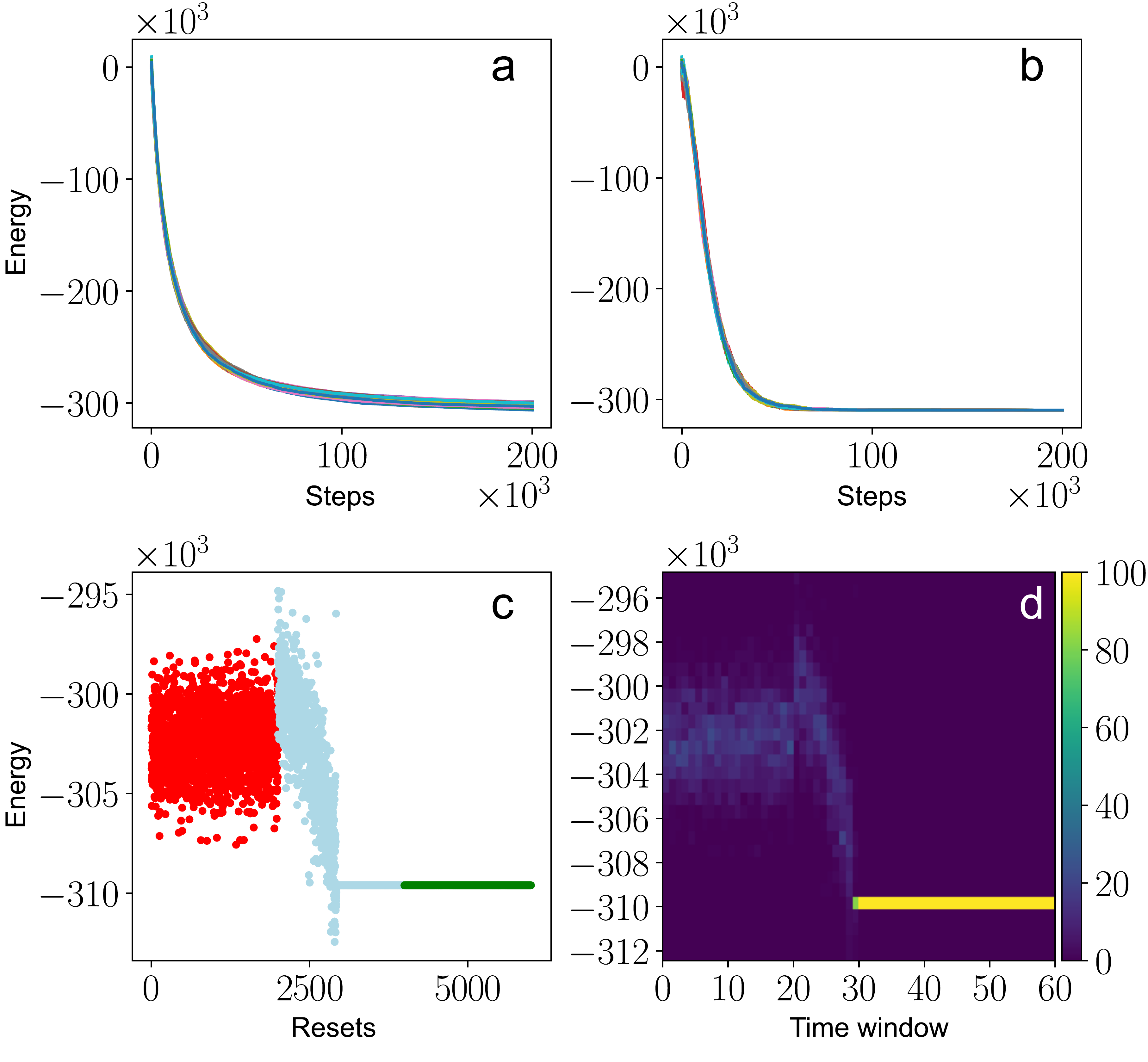}\caption{\label{fig:SO-N10k-1}Self-optimization simulation results for $N=10000$,
$\alpha=1\times10^{-9}$. (a),(b) Energy for 50 different initial
random states before and after learning, (c) Energy at the end of
convergence for a set without learning (resets 1--2000, red), during
learning (2001--4000, blue), and after learning (4001-6000, green).
(d) Histogram for the same energies in (c) averaged for each 100 resets. }
\end{figure}

As it can be seen from Fig.~\ref{fig:SO-N10k-1}, the behavior obtained
from the SO model for substantially enlarged systems is similar to
the small systems that were previously tested.

\section{Conclusions\label{sec:Conclusions}}

This work provides a novel implementation of a by-now classic model
of complex adaptive systems developed in the field of artificial life,
the SO model that reduces the memory access from $\mathcal{O}\left(N^{3}\right)$
to $\mathcal{O}\left(N^{2}\right)$. This substantial reduction in
computational effort enables the investigation of dramatically larger
systems. We demonstrated the applicability of the SO model to a network
that is two orders larger than was used in previous studies. Having
the possibility to compute large scale systems opens up a door to
new research investigations (e.g. more complex connectivity of the
weight matrix with several modular scales).

It is important to note that here we have updated the weights of the
system at every step of the simulation. Similar dynamics can be achieved
by making a weight update only at the reset of the system. This might
make the system less biologically plausible, but considerably faster,
opening the option to compute even larger systems than shown in this
work. 

It should be noted as well that although we have used a discrete state
in Algorithm~\ref{Alg-2}, it is relatively straightforward to generalize
the algorithm to a continuous state as well. In this case, the implementation
\ref{subsec:Integer-computation-of-W} in the Appendix would not be
applicable, and implementation \ref{subsec:Computation-of-energy}
would have to be modified slightly since $s_{x}^{2}\neq1$, but that
does not affect the salient point of this article, the improvement
of the algorithmic scaling.

As was mentioned in Sec.~\ref{sec:Introduction}, the SO model has
been demonstrated to be applicable for a wide range of complex adaptive
systems. With this new implementation, the future work in these various
fields may take a new turn into investigating systems with much higher
complexity. Researchers working on adaptive self-organization in neural,
gene regulatory, ecological, or social networks, who may have previously
overlooked this model because their networks operate at a much larger
scale, can now more easily apply the model to their systems of interest.
Given the robustness and generality of the model's self-optimization
process, we can expect that actual implementations of this process
are awaiting to be discovered in various natural systems. 

Looking further into the future, we might wish to address the network
topology studied. The current model of Hopfield neural network implies
a fully connected network of nodes. This is plausible or even necessary
to achieve a biologically relevant level of complexity. Building on
the substantial increase in scale of tractable SO problems demonstrated
in this article we may remove this requirement of fully connected
networks. Since we can now treat networks with modules that exceed
the size of entire systems previously treated, the modules themselves
can provide a sufficient level of complexity. Given the new source
of biological complexity, perhaps full connectivity between all modules
is no longer required. The number of connections (i.e. the size of
the populated weight matrix) determines the computational cost for
evaluating the model. Limiting the number of connections explicitly
accounted for leads to a less dramatic increase in computational cost
when further increasing the scale of the network with many more, but
partially interconnected, modules. Using an approach of this sort
we hope to study the behavior of interconnected systems on a truly
massive scale. 

\section{Data availability statement }

The model from the main text as well as the code used for the simulation
are available at \cite{weber_large-scale_2022}. 

\section{Acknowledgment }

We are grateful for the help and support provided by the Scientific
Computing and Data Analysis section of Research Support Division at
OIST.

\appendix
\label{Appendix}In addition to the main optimization outlined in
Section~\ref{sec:Implementation}, we performed a series of different
mathematical manipulations of the Hopfield update rule, \eqref{eq:state_update},
energy computation, \eqref{eq:E}, and Hebbian learning rule, \eqref{eq:w_update},
to further optimize the SO procedure to decrease its computational
time. These are used in \cite{weber_large-scale_2022} and described
below.

\subsection{Integer computation of W\label{subsec:Integer-computation-of-W}}

By dividing \eqref{eq:w_update} by $\alpha$, we can rewrite it to
\begin{equation}
\eta w_{ij}\left(t+1\right)=\eta w_{ij}\left(t\right)+s_{i}\left(t\right)s_{j}\left(t\right),\qquad\eta=\frac{1}{\alpha}.\label{eq:scaled_w_update}
\end{equation}
Assuming $\alpha$ is typically much smaller than one, $\eta$ can
be chosen to be a large integer number. We can therefore scale $\mathbf{W}$
by multiplying it by $\eta$ before the simulation starts. This in
turn allows us to set the type of $\mathbf{W}$ to 64-bit integer
and the type of $\mathrm{\mathbf{dW}}=\boldsymbol{\mathrm{s}}\boldsymbol{\mathrm{s^{\mathit{T}}}}$
to 8-bit integer, which decreases the computation time since floating-point
operations typically take longer to execute than integer operations.

\subsection{Computation of energy from the previous step\label{subsec:Computation-of-energy}}

Since at each step $t$ only one discrete state $s_{x}$ is updated
we can rewrite \eqref{eq:E} as
\begin{align}
E= & -\frac{1}{2}\left[\sum_{i\neq x}^{N}\sum_{j\neq x}^{N}w_{ij}s_{i}s_{j}+s_{i=x}\sum_{j\neq x}^{N}w_{xj}s_{j}\right.\nonumber \\
 & \left.+s_{j=x}\sum_{i\neq x}^{N}w_{ix}s_{i}+s_{i=x}w_{xx}s_{j=x}\right]\\
= & -\frac{1}{2}\left[\sum_{i\neq x}^{N}\sum_{j\neq x}^{N}w_{ij}s_{i}s_{j}+2s_{x}\sum_{i\neq x}^{N}w_{ix}s_{i}+w_{xx}\right],\label{eq:E_i=00003Dx}
\end{align}
since $s_{x}^{2}=1$ always, and $w_{xi}=w_{ix}$ since $\mathbf{W}$
is symmetric. Energy change between each step is
\begin{align}
\Delta E= & E\left(t-1\right)-E\left(t\right).\label{eq:dE}
\end{align}
Rearranging \eqref{eq:dE} and substituting \eqref{eq:E_i=00003Dx}
we get
\begin{align*}
E\left(t\right)= & E\left(t-1\right)+\\
 & \frac{1}{2}\left[\sum_{i\neq x}^{N}\sum_{j\neq x}^{N}w_{ij}s_{i}^{t-1}s_{j}^{t-1}+2s_{x}^{t-1}\sum_{i\neq x}^{N}w_{ix}s_{i}^{t-1}+w_{xx}\right.\\
 & \left.-\sum_{i\neq x}^{N}\sum_{j\neq x}^{N}w_{ij}s_{i}^{t}s_{j}^{t}-2s_{x}^{t}\sum_{i\neq x}^{N}w_{ix}s_{i}^{t}-w_{xx}\right]\\
= & E\left(t-1\right)+\left(s_{x}^{t-1}-s_{x}^{t}\right)\sum_{i\neq x}^{N}w_{ix}s_{i},
\end{align*}
where the double sum elements are canceled out since $s_{i\neq x}s_{j\neq x}$
do not change from $t-1$ to $t$, and therefore the time label from
them were removed for clearness. We can thus rewrite \eqref{eq:E}
as
\begin{equation}
E\left(t\right)=E\left(t-1\right)+\left(s_{x}^{t-1}-s_{x}^{t}\right)\left[\sum_{i}^{N}w_{ix}s_{i}^{t}-s_{x}^{t}w_{xx}\right].\label{eq:E2}
\end{equation}
With this optimization, we can compute the energy once using \eqref{eq:E}
and then compute energy every other step using \eqref{eq:E2}.

\subsection{Row and column update of W}

The straightforward computation \eqref{fig:W} can be slightly optimized
if we compute the entire matrix $\mathbf{dW}=\alpha\mathbf{s}\mathbf{s}^{{\rm T}}$
at the first step of each reset, and then just update one row and
one column in $\mathbf{dW}$ with $\mathbf{dW}\left[i,:\right]=\mathbf{dW}\left[:,i\right]=\alpha s_{i}\mathbf{s}$.
This can further be optimized by noting the fact that $\mathbf{W}$
is symmetric, so in principle, we only need to update the upper triangular
half of $\mathbf{dW}$ (and consequently $\mathbf{W}$) and then copy
that part to the lower triangular half at the end (as used in \cite{weber_large-scale_2022}).

\bibliographystyle{IEEEtran}
\bibliography{so-scaled-up-arXiv}

\end{document}